# Gated Si nanowires for large thermoelectric power factors

Neophytos Neophytou[1] and Hans Kosina[2]

[1]School of Engineering, University of Warwick, Coventry, CV4 7AL, UK

[2]Institute for Microelectronics, Vienna University of Technology, Gusshausstrasse 27-29/E360, Vienna, A-1040, Austria

N.Neophytou@warwick.ac.uk

# Abstract

We investigate the effect of electrostatic gating on the thermoelectric power factor of p-type Si nanowires (NWs) of up to 20nm in diameter in the [100], [110] and [111] crystallographic transport orientations. We use atomistic tight-binding simulations for the calculation of the NW electronic structure, coupled to linearized Boltzmann transport equation for the calculation of the thermoelectric coefficients. We show that gated NW structures can provide ~5x larger thermoelectric power factor compared to doped channels, attributed to their high hole phonon-limited mobility, as well as gating induced bandstructure modifications which further improve mobility. Despite the fact that gating shifts the charge carriers near the NW surface, surface roughness scattering is not strong enough to degrade the transport properties of the accumulated hole layer. The highest power factor is achieved for the [111] NW, followed by the [110], and finally by the [100] NW. As the NW diameter increases, the advantage of the gated channel is reduced. We show, however, that even at 20nm diameters, (the largest ones that we were able to simulate), a ~3x higher power factor for gated channels is observed. Our simulations suggest that the advantage of gating could still be present in NWs with diameters of up to ~40nm.

_Index terms:_ silicon nanowires, low-dimensional thermoelectrics, gated thermoelectrics, Boltzmann transport, thermoelectric power factor, Seebeck coefficient

Nanostructured and low-dimensional silicon based thermoelectric (TE) materials have attracted significant attention after they demonstrated large performance enhancement compared to the bulk material. The thermoelectric performance is quantified by the dimensionless figure of merit $ZT=\sigma S^2 T/\kappa$, where $\sigma$ is the electrical conductivity, $S$ is the Seebeck coefficient, and $\kappa$ is the thermal conductivity. Traditional thermoelectric materials suffer from low efficiency with $ZT$ around unity, attributed to the adverse interdependence of the electrical conductivity and Seebeck coefficient with carrier density, and high thermal conductivities. Nanomaterials such as 1D nanowires (NWs) [1, 2], 2D thin-layer superlattices [3, 4, 5, 6], as well as materials with embedded nanostructuring [7, 8, 9, 10, 11], however, demonstrated significant $ZT$ enhancements as a result of a large reduction in their thermal conductivity. Nanostructured Si, in particular, has demonstrated thermal conductivities as low as $\kappa$=1-2 W/mK (compared to $\kappa_{bulk}$=142 W/mK), which resulted in room temperature $ZT$~0.6 [1, 7, 10].

As the thermal conductivity approaches the amorphous limit it becomes evident that any additional improvements to the $ZT$ need to come from the power factor, which currently remains low. Techniques such as engineering the shape of the density-of-states function in low-dimensional materials [12, 13], or utilizing nanoscale potential barriers to enhance energy filtering [3, 4, 9, 14], in general improve the Seebeck coefficient. The electrical conductivity, however, undergoes a drastic reduction due to enhanced carrier scattering on the boundaries of the nanostructures. As a result, no significant improvements in the thermoelectric power factor $\sigma S^2$ were achieved.

Another approach towards achieving high power factors is the use of field effect modulation to achieve the high densities usually required (usually ~$10^{19}$-$10^{20}$/cm$^3$), rather than by direct doping, which is the commonly employed method. In heavily doped channels, impurity scattering from ionized dopants (IIS) dominates carrier scattering and mobility, and significantly limits the conductivity [15]. Therefore, modulation doping techniques [16, 17], or gated channels [18, 19, 20, 21, 22], that provide high carrier densities without direct doping could prove beneficial to the channel's thermoelectric properties. Such works have indeed demonstrated strong electric field modulation of the conductivity and thermopower [16, 17, 18, 19, 20, 21, 22].

The experimental works in Refs [18, 19, 20, 21, 22], however, consider thick structures, in which a high mobility transport layer is formed only along the structure's surfaces, whereas the volume of the channel remains mostly inert. In the best case, they report power factors similar to those of doped bulk materials [18]. It is, thus, difficult to examine whether gating could actually provide higher power factors, what would the magnitude of the improvement be, and at which channel dimension improvements can be achieved. In this work we theoretically address this point by performing an atomistic simulation study for the thermoelectric performance of gated versus doped Si nanowires.

We consider p-type Si NWs of diameters up to 20nm and three different transport orientations, the [100], [110], and [111]. The simulation approach for the gated structures is as follows [23]: i) The electronic structure of the NWs is calculated using the atomistic $sp^3d^5s^*$ tight-binding (TB) model. ii) The charge that resides in the bands and in the cross section of the NW is calculated assuming equilibrium statistics. iii) The 2D Poisson equation is solved in the cross section of the NW in order to obtain the electrostatic potential. A gated all around geometry is assumed with a $SiO_2$ oxide of thickness $t_{ox}$=1.1nm. Steps i)-iii) are iterated until self-consistency is achieved between charge and potential. This captures the influence of the gating field on the bandstructure. iv) Upon convergence the thermoelectric properties are extracted using the Linearized Boltzmann transport equation including all relevant scattering mechanisms, i.e. acoustic phonons, optical phonons, surface roughness scattering (SRS), and ionized impurity scattering (IIS) as described in detail in Refs [15, 23, 24]. IIS is only applied for doped/non-gated structures. For SRS we assume a 1D exponential autocorrelation function for the roughness, with roughness amplitude $\Delta_{rms}$ = 0.48nm and roughness correlation length $L_C$ =1.3nm. In the case of gated/undoped channels the SRS is computed using the radial electric field at the interface of the NW. For the ungated/doped channels, in which case we do not have a finite electric field, we compute SRS through the shift in the band edges of the NW with diameter modulation. In the case of doped structures the self-consistency is not necessary and only steps i) and iv) are employed. The methodology we employ, as described in Refs [15, 23, 24], is commonly employed to describe transport in low-dimensional semiconductors. We use appropriate assumptions and approximations such bulk phonon deformation potential scattering [24, 25, 26], uniformity in the doping profile and the resultant electrostatic potential, and rigid Si electronic bandstructures independent of doping. High doping and carrier concentrations could cause bandgap narrowing [27], but our results do not depend on the bandgap itself, rather on the position of the Fermi level with respect to the valence band.

Figure 1 shows the NW mobility versus carrier concentration in the cases of doped and gated NWs. Without loss in generality, we show results here for the [110] NW with diameter D=12nm. Four cases are depicted: i) Gated NW with phonon-limited scattering transport considerations (red-dot-solid line); ii) Gated NW with phonon scattering and SRS transport considerations (red-dot-dashed line); iii) Doped NW with phonon-limited scattering transport considerations (black-dashed line); iv) Doped NW with phonon scattering, SRS, and IIS transport considerations (black-solid line). In the gated NW cases the mobility is constant for low carrier concentrations, it increases by ~50% around densities $p=10^{19}/cm^3$, and then drops again. SRS degrades the phonon-limited mobility, but noticeably only at carrier concentrations above $10^{19}/cm^3$ (dashed versus solid red-dot lines). The reason why SRS is weak is that the electric field to achieved a hole *accumulation* layer is relatively weak (in contrast, the electric field to achieve an *inversion* layer is much stronger, as for example in transistor devices). The phonon-limited mobility for the doped structure is very similar to the phonon-limited mobility of the gated channel up to concentrations of $5x10^{17}/cm^3$. This result, of course, is just plotted for comparison purposes. Modulation in the carrier concentration can only be achieved either by gating or by doping, in which case IIS should additionally be included. The phonon-SRS-IIS-limited mobility for the doped channel (back-solid line), however, is strongly degraded as the carrier concentration increases. Note that SRS is very weak for doped NW channels of diameters above 10nm [15], and it is IIS that causes the observed mobility degradation. At carrier concentrations of $p=10^{19}/cm^3$, at which the power factor is usually maximized in thermoelectric materials, gating rather than doping could offer ~10x higher channel mobility, and could indeed, as we show below, provide superior thermoelectric power factors.

The mobility increase around concentrations p=$10^{19}$/cm$^3$ in the gated channels is attributed to modifications in the bandstructure of the NW as holes are electrostatically confined in an accumulation layer around the NW surface (see Inset of Fig. 2b) [23]. Figure 2a shows the bandstructure of the D=12nm [110] NW at low gate biases, with almost flat potential in its cross section. It also corresponds to the bandstructure of the doped NWs. Figure 2b shows the bandstructure of the gated channel under accumulation conditions, at carrier concentrations p~$2\times10^{19}$/cm$^3$. A noticeable difference in the bandstructures of Fig. 2a and Fig. 2b can be observed. The high energy region of the bandstructure in Fig. 2b, acquires a larger curvature. This is a consequence of the anisotropic shape of the heavy-hole band in common semiconductors. The heavy-hole band is warped, with 'wings' elongated in different directions, which makes them respond differently to different quantization conditions. For channels in the [110] transport direction under [1-10] electrostatic confinement, the bands that reside at lower energies are the ones with a heavy confinement mass, but light transport mass. A very similar situation is observed for the [111] channels also under [1-10] confinement [23]. Ultimately, this translates to the increase in the carrier mobility observed in Fig. 1. Figure 2c shows the effective mass of the D=12nm NWs versus gate bias for the [110] and [111] NWs. For the [111] NW the effective mass is even lower [15, 23, 28]. As the channel is driven into accumulation (increasing gate bias), the mass is drastically reduced. At very high gate biases the mass of the D=12nm NWs approaches the effective mass of the corresponding geometrically confined D=3nm NWs. These are depicted by the dots in the right side of Fig. 2c, which show atomistic calculations data from Refs [25, 28, 29]. As described in Ref. [28], the effective mass of the [110] and [111] NWs is also drastically reduced with diameter reduction, which indicates that physical and electrostatic confinement have a very similar effect on the bandstructure. Note that the effective mass values in Fig. 2c are calculated by extracting the average injection velocity $\left\langle v_x^+ \right\rangle$ of the positive going states under non-degenerate conditions, and extract the mass by $\left\langle m_{eff}^* \right\rangle = 2k_B T / \pi \left\langle v_x^+ \right\rangle^2$, as described in Ref. [28]. We use this method because it also includes information from the overall bandstructure, which is composed of non-parabolic bands of various curvatures, and provides a better 'estimate' for an 'overall' mass.

Therefore, due to: i) the absence of ionized impurity scattering, ii) the weak influence of SRS, and iii) the increase in the curvature of the bands, the electrical conductivity of the gated channel largely surpasses that of the doped channel. This is shown in Fig. 3a which depicts the conductivity versus carrier density for the same four channel cases as shown above in Fig. 1. The conductivity of the gated channel (red lines) is the highest. Including SRS on top of phonon scattering affects the conductivity only to a small degree, and only at concentrations above $p=5x10^{19}/cm^3$. The phonon-limited conductivity of the channel with flat potential in its cross section (black-dashed line) is somewhat lower compared to that of the gated structures. Once IIS (and SRS) is additionally considered in the calculations (as in a doped channel), however, the conductivity significantly drops (black-solid line). The Seebeck coefficient, on the other hand, improves when scattering becomes stronger as shown in Fig. 3b. In general, additional scattering mechanisms improve the Seebeck coefficient since this quantity follows the inverse trend compared to conductivity. Adding SRS to the phonon-limited result of the gated channel improves the Seebeck coefficient at higher carrier concentrations (red-dot lines). Similarly, ionized impurity scattering in the case of the doped channel (solid-black line), largely improves the Seebeck coefficient. That increase, however, is not large enough to compensate for the large degradation of the electrical conductivity caused by ionized impurity scattering. The thermoelectric power factor shown in Fig. 3c is much larger in the case of the gated channel compared to the doped channel. The power factor of the gated channel peaks at concentrations around $p\sim10^{19}/cm^3$, and it is ~5x higher than that of the doped channel. This qualitatively and quantitatively demonstrates the advantage of the gated NW channels for achieving large thermoelectric power factors compared to the traditionally used doped materials. It is interesting to observe that SRS, which could degrade the conductivity of the gated channel, becomes important at concentrations much larger than the optimal ones needed for high power factors, $p\sim10^{19}/cm^3$. On the other hand, IIS is already strong enough at this concentration to cause significant degradation to the power factor of the doped channel. Similar qualitative conclusions were also shown for thin layers, however in that case the improvement due to gating was somewhat weaker [30].

A very similar behavior is observed for the [111] NW, which has a larger power factor as well [24]. In general, the performance is highly anisotropic. Figure 4 shows the thermoelectric power factor versus carrier concentration for NWs in the [100] (blue lines), [110] (red lines) and [111] (green lines) transport orientations. For each orientation, we compare the power factor between the gated structure (including phonon scattering and SRS in the calculation), and the doped channel (including phonon scattering, SRS, and IIS in the calculation). The [111] NW has the highest power factor, a factor of ~2x higher than the [110] NW, and almost ~3x higher than that of the [100] NW. Importantly, in all three NW cases gating largely improves the thermoelectric power factor by a similar amount, approximately a factor of ~5x. Note that this anisotropy dependence, is actually a direct correspondence of the anisotropy in mobility of the p-type NWs as described in Refs [15, 28, 31], whereas the difference in the Seebeck coefficient of the NWs is not significant.

As we have discussed, the beneficial effect of channel gating over doping originates from the high mobility of the carriers that reside in the accumulation layer around the NW surface. The volume in the core center of the NW also contributes to the power factor when a significant amount of charge resides there, especially for narrow NWs. As the NW diameter is increased, however, less charge density accumulates in the core of the NW because the gate field is screened by the surface charge. The core of the NW then participates less to transport and for very large diameters it could even become inactive. In that case the overall conductivity, Seebeck coefficient, and power factor will be the weighted average of these quantities in the hole accumulation layer and in the core volume of the NW, with weighting factor being the area that each region occupies in the cross section of the NW. As the diameter increases, the core area increases, the power factor decreases, and the advantage of gating is lost.

Figure 5 demonstrates this by showing the power factor versus carrier density for gated [110] NWs with diameters increasing from D=12nm up to D=20nm. Phonon scattering and SRS are considered in the calculation. For comparison we also show the power factor of the doped [110] D=12nm NW as in Fig. 3c. A decline in the power factor of the gated NWs is observed as the diameter increases. For NW diameters up to 20nm gating is still beneficial, but the maximum power factor is now only ~3x higher than that of the doped channel (compared to ~5x in the D=12nm NW case). Extrapolating the trend of Fig. 5, however, we estimate that the power factor advantage for the gated channels could be retained even up to NWs with diameters D~40nm. (Note that these self-consistent atomistic calculations are numerically very expensive. Simulating atomistically NWs with diameters up to D=40nm is computationally prohibitive).

The fact that the benefits of gating appear at channel dimensions below a certain size, could partially explain why measurements of the thermoelectric power factor in gated structures did not demonstrate values significantly larger than those of the doped bulk material. These measurements are usually performed in thick gated structures, and despite the observation of strong gate modulation of the conductivity and Seebeck coefficient, the power factor improvements were lower than what we predict [18, 19, 20, 22]. Here, however, we theoretically show that field-effect modulation can lead to large performance improvements. In another recent work, Curtin et al. [21] indicated that gating can also improve the power factor of n-type [100] NWs. This complements our work, as thermoelectric modules utilize both n- and p-type materials.

In conclusion, we have investigated the effect of gating in the thermoelectric power factor of narrow p-type Si NWs of diameters up to 20nm using self-consistent atomistic tight-binding calculations and Boltzmann transport. We show that for NWs with diameter D=12nm, gating rather than doping can result in a 5-fold improvement in their power factor. This improvement, however, is reduced as the NW diameter is increased. For NWs with diameters up to D=20nm the advantage is reduced to ~3x. We estimate that some degree of power factor improvement with gate field rather than doping could still be retained for NWs with diameters even up to ~40nm. We finally show that the NW orientation is also an important design parameter, with the p-type [111] NW performing ~2x higher than the [110] NW, and ~3x higher than the [100] NW.

*Acknowledgement:* The work leading to these results has received funding from the European Community's Seventh Framework Programme under grant agreement no. FP7-263306.

Figure 1:

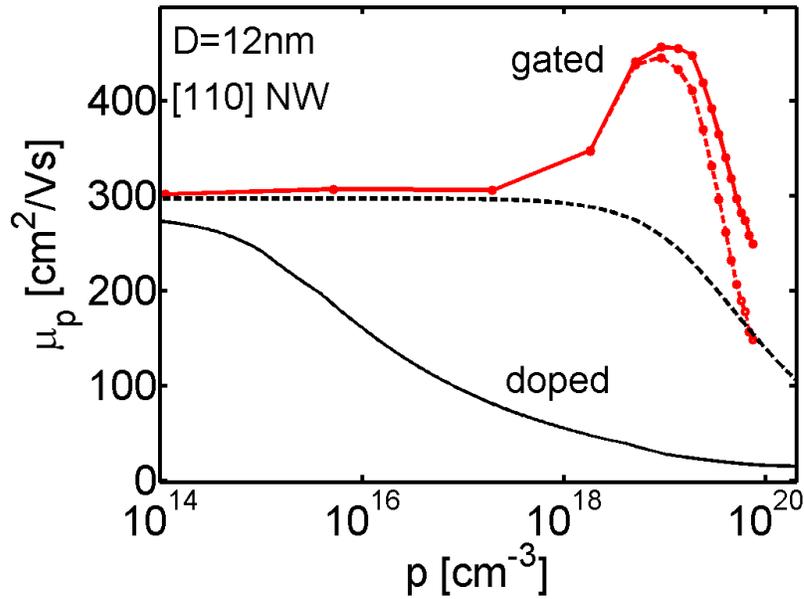

## Figure 2:

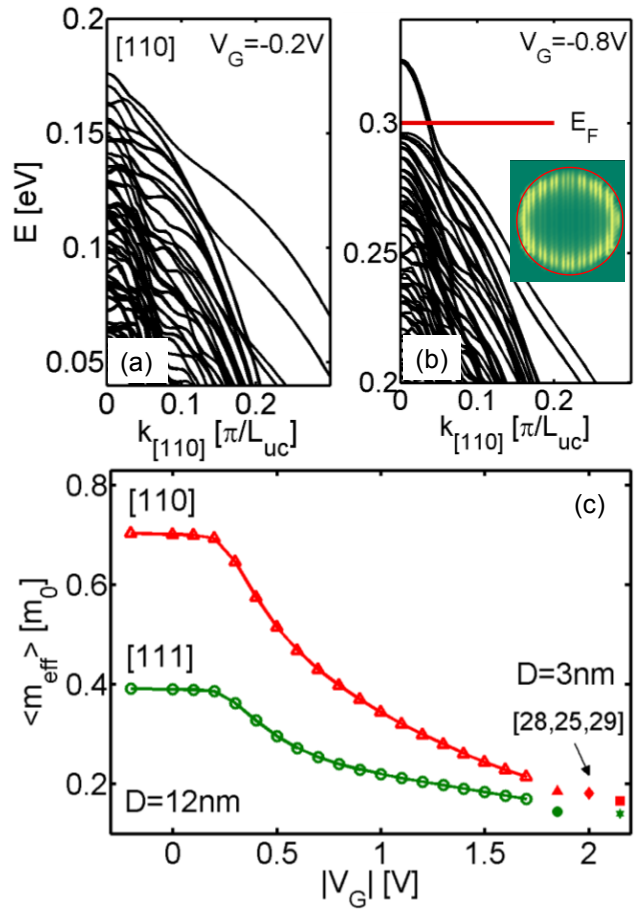

## Figure 2 caption:

The electronic bandstructure of the gated D=12nm, [110] NW. (a) Low gate bias, with almost flat potential in the cross section of the NW. (b) High gate bias, in which case a potential well is formed around the NW surface. The Fermi level in (b), is indicated by the red line. Inset of (b): The charge density in the cross section of the NW. A hole accumulation layer is shown in bright-yellow. Green-dark color indicates low charge carrier concentration regions. (c) An 'estimate' of the effective mass of the D=12nm NWs in the [111] and [110] orientation versus gate bias. The dots at the left of the figure are calculations for the effective mass of the D=3nm [111] (green, lower dots) and [110] (red, upper dots) NWs from Refs. [28, 25, 29].

Figure 3:

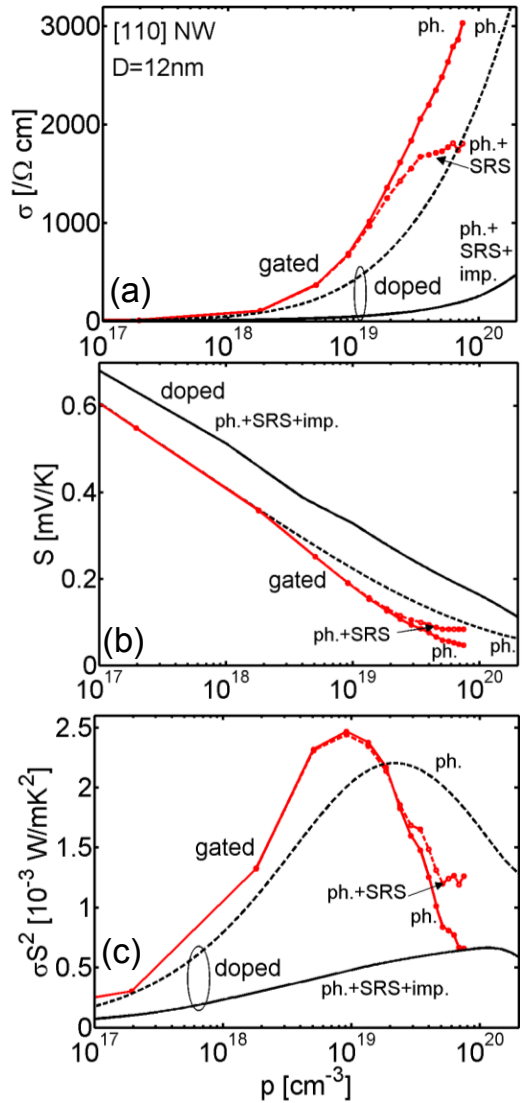

The thermoelectric coefficients of the D=12nm [110] NW versus carrier density. (a) The electrical conductivity. (b) The Seebeck coefficient. (c) The power factor. Four different channel situations are shown: (i) Gated NW under phonon scattering-limited considerations (red-dot-solid lines). (ii) Gated NW under phonon scattering and SRS considerations (red-dot-dashed lines). (iii) Doped (non-gated) NW under phonon scattering-limited considerations (black-dashed lines). (iv) Doped (non-gated) NW under phonon scattering, IIS, and SRS considerations (black-solid lines).

Figure 4:

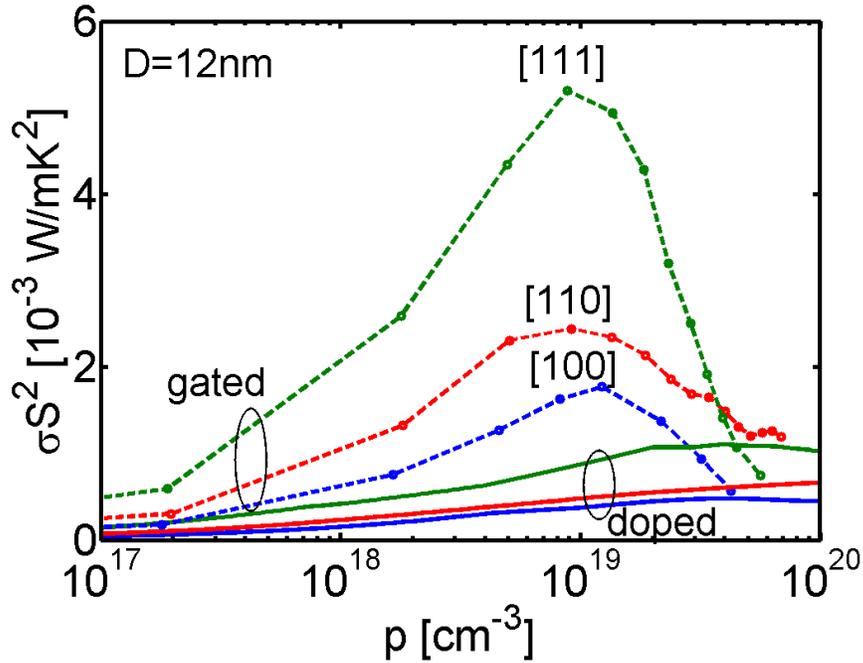

Figure 4 caption:

The thermoelectric power factor of the D=12nm NWs versus carrier density. Results for the [100] (blue lines), [110] (red lines) and [111] (green lines) NWs are shown. For each NW, two different channel situations are shown: (i) Gated, undoped NWs, under phonon scattering and SRS considerations (dashed-dot lines). (ii) Doped, non-gated NWs under phonon scattering, IIS, and SRS considerations (solid lines).

Figure 5:

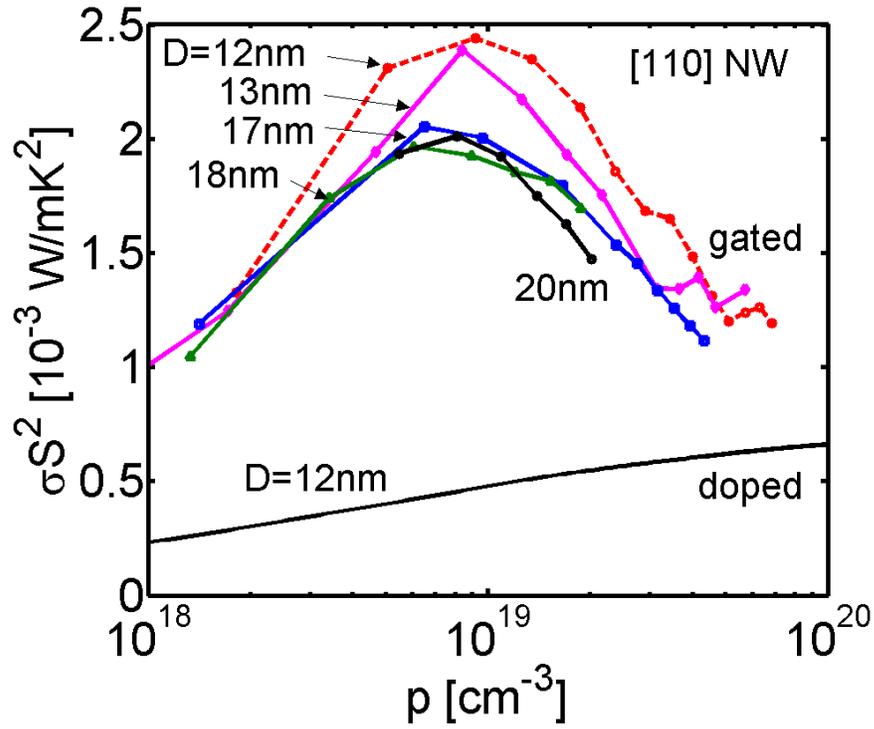

Figure 5 caption:

The thermoelectric power factor of gated [110] NWs with diameters D=12nm, 13nm, 17nm, 18nm, and 20nm as indicated, versus carrier density. Phonon scattering and SRS are considered. The power factor for the doped case (as in Fig. 3c) is also depicted by the solid-black line. In this case, phonon scattering, IIS, and SRS are considered.